\documentclass[a4paper,11pt]{article}

\usepackage{jheppub} 

\usepackage[T1]{fontenc} 

\newcommand{\eref}[1]{eq.~(\ref{#1})}

\title{\boldmath Gradient Flow of O(N) nonlinear sigma model at large N}

\author[a]{Sinya Aoki,}
\author[a,1]{Kengo Kikuchi,\note{Corresponding author.}}
\author[b]{Tetsuya Onogi}

\affiliation[a]{Yukawa Institute for Theoretical Physics, Kyoto University, Kyoto 606-8502, Japan}
\affiliation[b]{Department of Physics, Osaka University, Toyonaka, Osaka 560-0043, Japan}

\emailAdd{saoki@yukawa.kyoto-u.ac.jp}
\emailAdd{kengo@yukawa.kyoto-u.ac.jp}
\emailAdd{onogi@phys.sci.osaka-u.ac.jp}

\abstract{We study the gradient flow equation for the O(N) nonlinear sigma model in two dimensions at large $N$.
We parameterize solution of the field at flow time $t$ in powers of bare fields by introducing 
the coefficient function $X_n$ for the $n$-th power term ($n=1,3,\cdots$).
Reducing the flow equation by keeping only the contributions  at leading order in large $N$, 
we obtain a set of equations for $X_n$'s,  which can be solved iteratively starting from $n=1$. 
For $n=1$ case, we find an explicit form of the exact solution. 
Using this solution, we show that the two point function at finite flow time $t$
is finite.  As an application, we obtain the non-perturbative running coupling defined from 
the energy density. We also discuss the solution for $n=3$ case. }

\begin{document} 
\hspace{10cm}YITP-14-109, OU-HET-847
\maketitle
\flushbottom

\newpage 
\section{Introduction}
In recent years, the gradient flow equation \cite{Luscher:2010iy} has been the focus of attention.
The gradient flow equation is originally proposed in the context of  the SU(N) lattice gauge theory \cite{Luscher:2009eq} 
and the SU(N) Yang-Mills theory \cite{Luscher:2010iy}. In Ref.~\cite{Luscher:2013cpa}, L\"{u}sher also gave the matter fields version of the gradient flow equation. 
The gradient flow can also be viewed as a nice way of smearing the bare field respecting the gauge symmetry 
which could tame fluctuations of the operator arising from the contributions at high momentum scale. 
For this reason, the gradient flow can give a useful physical quantities which are numerically very 
stable and well-defined in the continuum. 

With these remarkable features, various applications 
of the gradient flow have emerged \cite{Luscher:2013vga};  scale-setting and/or running \cite{Luscher:2010iy, Fodor:2012qh, Borsanyi:2012zs, Hasenfratz:2014rna, Fritzsch:2013je, Fritzsch:2013hda, Ramos:2013gda, Rantaharju:2013bva, Fritzsch:2013yxa, Fodor:2012td, Fodor:2014cpa}, chiral condensate \cite{Luscher:2013cpa}, topological charge \cite{Luscher:2010iy}, 
and operator renormalization from small $t$ behavior \cite{DelDebbio:2013zaa, Suzuki:2013gza, Makino:2014wca, Shindler:2013bia, Monahan:2013lwa}, and other applications \cite{Asakawa:2013laa, Bar:2013ora, Brida:2013mva}. 
In view of these successes, it is worth extending the method to not only gauge theory but also other quantum field theories.

In a recent paper \cite{Kikuchi:2014rla}, the generalization of the gradient flow equation for field theory with 
non-linearly realized symmetry are proposed. This equation gives a unified method to construct a gradient flow equation of the action with non-linearly realized symmetry, for example, the supersymmetric Yang-Mills theory, the O(N) nonlinear sigma model in two dimensions. Of course the equation can reconstruct the equation of the Yang-Mills theory and the lattice gauge theory.    

In this paper, we focus on this O(N) nonlinear sigma model in two dimensions \cite{Polyakov:1975rr, Migdal:1975zf, Balog:2012db}. Since this theory is known 
to be a good toy model of the Yang-Mills theory with asymptotic freedom and non-perturbative generation of the mass gap, and 
exactly solvable, it can be an ideal laboratory for the theoretical study of the gradient flow. Recently, using the gradient flow method, the ultraviolet finiteness of the O(N) nonlinear sigma model in two dimensions was proved to all order in perturbation theory \cite{Makino:2014sta}.

The most interesting point to study the O(N) nonlinear sigma 
model in two dimensions is that the model is solvable at large $N$ limit \cite{'tHooft:1973jz}.  
Therefore, one could also expect that the finiteness proof may be possible at the non-perturbative level,
while such a non-perturbative proof of the finiteness for correlation function of the operators constructed from the solution to the gradient flow equation seems difficult for the Yang-Mills theory or QCD, despite its importance, since  these theories are not exactly solvable.
It would therefore be important to give a non-perturbative proof of finiteness and also carry out various applications in an exactly solvable model in order to get a deeper theoretical insight. 

In this paper, we study the finiteness of the solution to the gradient flow equation in the O(N) nonlinear sigma model in 
two dimensions  at large $N$. Due to the interaction terms in the flow equation,  the single scalar field solution to the gradient 
flow equation is given by the infinite sum of the convolutions in  $n$-th order multiple bare fields, where $n=1,3, \cdots$.  
We show that at large $N$ after dropping the subleading contributions a drastic reduction takes place and one obtains 
a closed set of equations for $n=1, 3, \cdots$ ,  which can in principle be solved iteratively.  In particular, we give an 
explicit solution for $n=1$, from which we can construct  the exact expression of the two point function at finite flow time $t$.
From the exact expression, one can show that the two point function is finite at finite $t$. We also give a formal solution to the $n=3$ case, from which one can obtain the connected four point function.

In Section~\ref{sec:model},  we introduce the O(N) nonlinear sigma model in two dimensions and solve the gap equation to determine the vacuum in the large $N$ limit. 
In Section~\ref{sec:gflow},  we introduce the gradient flow equation of this model, and solve it for $n=1$ in Section~\ref{sec:solution}.
The finiteness of the two point function for nonzero flow time is shown in Section~\ref{sec:finiteness}.
As an application, the non-perturbative running coupling is discussed in Section~\ref{sec:application}.
In Section~\ref{sec:discussions}, the four point function for nonzero flow time is briefly considered, though the discussion on the finiteness is left to future studies. We summarize our results of this paper in Section~\ref{sec:summary}.
In Appendix~\ref{app:4-pt}, the four point function in the two dimensional model is calculated,
and the solution to the gradient flow equation for $n=3$ is presented in Appendix~\ref{app:sol-4pt}. 
We also give an alternative way of solving the gradient flow equation using the Schwinger-Dyson equation 
of the two and four point functions in Appendix~\ref{app:alternative}.

\newcommand{\nn}{\nonumber}
\def\dfrac#1#2{\displaystyle\frac{#1}{#2}}
\newcommand{\ovl}[1]{\overline{#1}}
\newcommand{\wt}[1]{\widetilde{#1}}
\newcommand{\eq}[1]{Eq.~(\ref{#1})}
\newcommand{\eqn}[1]{(\ref{#1})}
\newcommand{\p}{\partial}
\newcommand{\pslash}{p\kern-1ex /}
\newcommand{\qslash}{q\kern-1ex /}
\newcommand{\lslash}{l\kern-1ex /}
\newcommand{\sslash}{s\kern-1ex /}
\newcommand{\kaslash}{k_a\kern-2ex /}
\newcommand{\kbslash}{k_b\kern-2ex /}
\newcommand{\Dslash}{{\cal D}\kern-1.5ex /}
\newcommand{\bpsi}{\overline{\psi}}
\newcommand{\bc}{\overline{c}}
\newcommand{\tr}{{\rm tr}}
\newcommand{\vev}[1]{\langle #1 \rangle}
\newcommand{\VEV}[1]{\left\langle{\rm T} #1\right\rangle}
\newcommand{\beqa}{\begin{eqnarray}}
\newcommand{\eeqa}{\end{eqnarray}}

\section{O(N) nonlinear sigma model}
\label{sec:model}
\subsection{Model and the gap equation}
We consider the O(N) nonlinear sigma model in two dimensional Euclidean space. The generating functional with source $J$ is given as
\begin{eqnarray}
Z(J) =\int {\cal D}\alpha(x)  {\cal D}\varphi \exp\left[- S +  \int d^2 x \left\{ i\alpha(x) \left(  \varphi(x)\cdot\varphi(x) - 1 \right) + J(x)\cdot \varphi(x) \right\}\right], 
\end{eqnarray}
where $\varphi_\alpha (\alpha=1,\cdots, N) $ is the scalar fields in the vector representation of O(N)  with unit length, whose action $S$ is given by
\begin{eqnarray}
S = \frac{1}{2g^2}\int d^2 x\,  \partial^\mu \varphi\cdot \partial_\mu \varphi ,
\end{eqnarray}
and the inner product is understood as $A\cdot B = \sum_{\alpha=1}^N A^\alpha B^\alpha$.

After integrating $\varphi$ field, we obtain
\begin{eqnarray}
Z(J) &=&\int {\cal D}\alpha(x) e^{-S_{\rm eff} (\beta, J)} \\
S_{\rm eff} (\beta, J) &=& 
N\left(i \int d^2 x \beta(x)  +\frac{1}{2} \mbox{tr} \ln K \right) 
 - \frac{\lambda}{2N}   J^\alpha(x) K^{-1}(x,y)J^\alpha(y) 
\\
K(x,y) &=& \left[-\square-2i\lambda \beta(x)\right]\delta^{(2)}(x-y).
\end{eqnarray}
where we have defined the rescaled field  $\beta(x)$ and rescaled coupling $\lambda$ as 
\begin{eqnarray}
\beta(x) \equiv \alpha(x)/N,  & \lambda \equiv g^2 N .
\end{eqnarray}
In this paper, according to the context,  the repeated coordinate index is understood as its integration such that
\beqa
F(x) G(x) &=& \int d^2 x\, F(x) G(x).
\eeqa

In the large $N$ limit, the path integral over $\beta$ is dominated by the stationary point determined by the following gap equation.
\begin{eqnarray}
1 
  &=& \lambda \int \frac{d^2 p}{(2\pi)^2} \frac{1}{p^2+m^2}
\label{eq:Gap}
\end{eqnarray}
where $m^2= -2i \lambda \langle \beta \rangle$ with $\langle \beta \rangle $ being the vacuum expectation 
value of $\beta$.
Introducing momentum cutoff $\Lambda$, the solution is given as
\begin{eqnarray}
1 = \frac{\lambda}{4\pi} \ln \frac{\Lambda^2+m^2}{m^2}.
\end{eqnarray}
Since $m$ is the nonperturbative physical mass of the scalar field, we impose the renormalization condition that
\begin{eqnarray}
m^2 = \mbox{finite}, 
\end{eqnarray}
which implies that the coupling $\lambda$ vanishes in the $\Lambda\rightarrow\infty$ limit as
\begin{eqnarray} 
\lambda = \frac{4\pi}{\ln \dfrac{\Lambda^2+m^2}{m^2}}.
\end{eqnarray} 
This shows that the theory is asymptotically free. 

\subsection{Power-counting in the large N expansion}

We expand $S_{\rm eff}$  around the stationary point $\langle \beta \rangle$ as
\beqa
S_{\rm eff}(\langle\beta\rangle +\beta, J) &=& \frac{N}{2}\beta(x) D_0(x,y) \beta(y)+N \sum_{n=3}^\infty V_n(\{z\}_n) \prod_{i=1}^n \beta(z_i) \nn \\
&+& \frac{1 }{N}\sum_{n=0}^\infty J^\alpha (x)  T_n (x,y,\{z\}_n) J^\alpha(y) \prod_{i=1}^n \beta(z_i) ,
\eeqa
where $\{z\}_n =z_1,\cdots,z_n$ and the first few terms are given by
\beqa
D_0(x,y) &=& 2 \lambda^2 \left[K_0^{-1}(x,y)\right]^2, \qquad
K^{-1}_0(x,y) = \int \frac{d^2p}{(2\pi)^2}\frac{ e^{i p(x-y)}}{p^2+m^2}, \\
 T_0 (x,y)  &=&-\frac{\lambda}{2}  K_0^{-1}(x,y) , \\
 T_1(x,y,z_1) & =& -i\lambda^2  K_0^{-1}(x,z_1)  K_0^{-1}(y,z_1)  , \\
T_2(x,y,z_1,z_2)  &=&   2 \lambda^3  K_0^{-1}(x,z_1) K_0^{-1}(z_1,z_2) K_0^{-1}(z_2,y) .
\eeqa

Now let us  consider the power counting in the large $N$ expansion.
We define
\beqa
F(J) &\equiv& \log Z(J)= \sum_{k=0}^\infty F_{2k} J^{2k}
\eeqa
where $F_{2k}$ corresponds to the connected $2k$-pt function. 
Since the propagator, $(ND_0)^{-1}$, is $O(1/N)$, a diagram which contains $v_n$ vertices of the type
$V_n$, $t_n$ vertices of the type $T_n$ and $I$ internal propagators, behaves as $N^{\nu}$ , where 
\beqa
\nu &=& \sum_{n=3}^\infty v_n -\sum_{n=0}^\infty t_n - I=  \sum_{n=3}^\infty v_n(1-n/2) -\sum_{n=0}^\infty t_n (1+n/2),
\eeqa
while the number of $J^{2k}$ is given by
\beqa
k= \sum_n t_n .
\eeqa
Therefore the leading power of $N$ for $F_{2k}$, denoted $N^{\nu_{2k}}$,  is given by
\beqa
\nu_{2k} &=& \left\{
\begin{array}{ccc}
-1, & t_0=1, & k=1 \\
-3, & t_1=2, & k=2 \\
-(2k-1), & t_1=k, v_k=1, & k\ge 3 \\
\end{array}
\right. ,
\eeqa
which corresponds to the tree level diagrams.

For $k=1$ and $2$, for example, we have
\beqa
F_2(x,y) &=&-\frac{1}{N}T_0(x,y)= \frac{\lambda}{2N} K_0^{-1}(x,y),  \label{eq:F2}\\
F_{4}(x_1,x_2,x_3,x_4) &=& \frac{1}{2N^3} T_1(x_1,x_2,z_1) D_0^{-1}(z_1,z_2) T_1(x_3,x_4,z_2) +O\left(N^{-4}\right).
\label{eq:F4}
\eeqa
Thus the 2-pt function is $O(1/N)$ and is given by
\beqa
\langle \varphi^\alpha (x) \varphi^\beta(y) \rangle &=&  \delta^{\alpha\beta}\frac{\lambda}{N} K_0^{-1}(x-y)=
 \delta^{\alpha\beta}\frac{\lambda}{N}\int \frac{d^2p}{(2\pi)^2} \frac{e^{ip(x-y)}}{p^2+m^2}.
 \label{eq:2pt}
\eeqa
at the leading order of  the large $N$ expansion.

\section{Gradient flow equation} 
\label{sec:gflow}
The gradient flow equation of the $O(N)$ nonlinear sigma model is defined for the field 
$\phi^\alpha(t,x)$, where an additional parameter $t$  corresponds to the flow time, with an initial condition that
$\phi^\alpha(0,x)=\varphi^\alpha(x)$. 
Since the field  $\varphi^\alpha(x)$ is subject to the constraint
$\sum_{\alpha=1}^N (\varphi^\alpha)^2=1$,  we impose the same constraint for $\phi^\alpha(t,x)$, so that 
 the $N$-th component can be expressed as 
 \begin{eqnarray}
 \phi^N = \pm \sqrt{1-\sum_{a=1}^{N-1} (\phi^a)^2}.
 \label{eq:constraint}
\end{eqnarray}
Substituting eq.~(\ref{eq:constraint}), the action can be rewritten as
\begin{eqnarray}
S=\frac{1}{2g^2} \int d^2x \sum_{a,b=1}^{N-1}g_{ab}(\phi)\left(\partial_\mu \phi^a \partial_\mu\phi^b \right)
\end{eqnarray}
where $a=1,2,\cdots,N-1$. Here, the metric for the $O(N)$ nonlinear sigma model is given by
\begin{eqnarray}
g_{ab}(\phi) = \delta_{ab} + \frac{\phi^a \phi^b}{1-\displaystyle{ (\phi^c)^2}}, &
g^{ab}(\phi)=\delta_{ab}-\phi^a\phi^b.
\end{eqnarray}
In Ref. \cite{Kikuchi:2014rla}, it was shown that the gradient flow equation for the field theory with nontrivial 
metric in the field space is given by
\begin{eqnarray}
\frac{d}{dt}{\phi}^a(t,x)=- g^{ab}(\phi(t,x))\frac{\delta  S}{\delta\phi^b(t,x)}.
\end{eqnarray}

We then obtain the gradient flow equation of the O($N$) nonlinear sigma model in two dimensions as
\begin{eqnarray}
\frac{d}{dt}\phi^a=\square\phi^a+\phi^a\partial_\mu\vec{\phi}\cdot\partial_\mu\vec{\phi}+\frac{\phi^a (\partial_\mu \vec{\phi}^2)^2}{4(1-\vec{\phi}^2)}.\label{eq:Gflow}
\end{eqnarray}
Here we rescaled  as $t\rightarrow g^2 t$ and the following notation for the summation over the indices are introduced
\begin{eqnarray}
 \vec{\phi}^2 = \sum_{b=1}^{N-1} (\phi^b)^2, & &
(\partial_\mu \vec{\phi})^2 = \displaystyle{\sum_{b=1}^{N-1}}(\partial_\mu \phi^b)^2 .
\end{eqnarray}


 \section{Solution to the gradient flow equation in the large N expansion}
 \label{sec:solution}
 In this section, we propose a method to solve the gradient flow equation non-perturbatively in the large $N$ expansion, and explicitly give a non-perturbative solution needed for the two point function of the $\phi$ field. 
 
 \subsection{Ansatz for the solution}
For the solution to the gradient flow equation, we take the following form   
\beqa
\phi^a(t,p) &=& f(t) e^{-p^2 t} \sum_{n=0}^\infty : X_{2n+1}^a(\varphi, p, t) :
\label{eq:sol}
\eeqa 
where $X_{2n+1}$ only contains $2n+1$-th order of $\varphi$, 
and $: {\cal O}: $ represents the "normal ordering", where self-contractions within  the operator ${\cal O}$ are prohibited. Formally we can define the normal ordering recursively in the perturbation theory around the large $N$ vacuum as
\beqa
:\varphi^a(p): &=& \varphi^a(p) \\
\langle :\varphi^{a_1}(p_1) \varphi^{a_2}(p_2): {\cal O} \rangle &=& \langle \varphi^{a_1}(p_1) \varphi^{a_2}(p_2) {\cal O}\rangle - \langle \varphi^{a_1}(p_1) \varphi^{a_2}(p_2)\rangle \langle {\cal O}\rangle\\
\langle : \prod_{j=1}^n \varphi^{a_j}(p_j) : {\cal O} \rangle &=& \langle  \prod_{j=1}^n \varphi^{a_j}(p_j)  {\cal O} \rangle
-\sum_{k\not= l}^n \langle \varphi^{a_k}(p_k) \varphi^{a_l}(p_l)\rangle  \langle  : \prod_{j\not=k,l}^{n-2} \varphi^{a_j}(p_j):  {\cal O} \rangle
\eeqa
for an arbitrary operator ${\cal O}$.
From the initial condition for $\varphi$, we have 
\beqa
X_1^a(\varphi,p,t) &=& \varphi^a(p), \qquad f(0)=1, \qquad X_{2n+1}^a(\varphi,p,0) = 0, \ n \ge 1 .
\eeqa

The gradient  flow equation in the momentum space is written as
\beqa
L^a(t,p) &\equiv & \dot\phi^a(t,p) + p^2 \phi^a(t,p) \nn \\
=R^a(t,p)&\equiv& - \int_{p}^{3}  \phi^a(t,p_1) (p_2\cdot p_3) \vec\phi(t,p_2)\cdot\vec\phi(t,p_3)
- \sum_{n=0}^\infty \int_{p}^{ 2n+5}  \phi^a(t,p_1) \nn \\
&\times & \frac{p_2+p_3}{2}
\cdot\frac{p_4+p_5}{2}
\prod_{j=1}^{n+2} \vec\phi(t,p_{2j})\cdot\vec\phi(t,p_{2j+1}) ,
\eeqa 
where we define
\beqa
\int_p^n &\equiv& \prod_{i=1}^n \int \frac{d^2 p_i}{(2\pi)^2} \hat\delta\left(\sum_{i=1}^n p_i - p\right),  \quad
\hat\delta(p) \equiv (2\pi)^2\delta^{(2)}(p) .
\eeqa
The left hand side can be expressed in term of the solution eq.~(\ref{eq:sol}) as
\beqa
L^a(t,p) &=& e^{-p^2t} \left[  \dot f(t) \sum_{n=0}^\infty :X_{2n+1}^a(\varphi, p,t) : +f(t) \sum_{n=1}^\infty : \dot X_{2n+1}^a(\varphi, p,t) : \right] .
\eeqa

In the present approach,  we are looking for the solution of the field $\phi(t,p)$ itself. As an alternative approach, one could 
also solve the $2n$-point correlation function $\langle \prod_{i=1}^{2n} \phi^{a_i}(t_i, p_i) \rangle$. This will be given 
in Appedix \ref{app:alternative}.

\subsection{Solution for $O_1$}
Taking ${\cal O}_1$ as the order $\varphi$ operator, we evaluate $L^a$ and $R^a$ at the leading order of the large $N$ expansion as
\beqa
\langle L^a(t,p) {\cal O}_1 \rangle &=&  e^{-p^2 t} \dot f(t) \langle \varphi^a(p) {\cal O}_1 \rangle, \\
\langle R^a(t,p) {\cal O}_1 \rangle &=& \lambda e^{-p^2 t} f^3(t) I(t) \langle \varphi^a(p) {\cal O}_1 \rangle + O(1/N)
\eeqa
where 
\beqa
I(t) &=& \int \frac{d^2q}{(2\pi)^2}\frac{q^2}{q^2+m^2} e^{-2q^2 t} .
\eeqa

The gradient flow equation that $ \langle  L^a(t,p) {\cal O}_1 \rangle = \langle R^a(t,p) {\cal O}_1 \rangle $ implies
\beqa
\dot f(t) &=& \lambda f^3(t) I(t), 
\eeqa
which can easily  be solved as
\beqa
f(t) &=& [ 1- 2 \lambda J(t) ]^{-1/2},
\eeqa
where
\beqa
J(t) &=& \int_0^t d s I(s) = \frac{1}{8\pi}\left[ \log \frac{\Lambda^2+m^2}{m^2} -\int_0^{\Lambda^2} d x \frac{e^{-2x t}}{x+m^2} \right] \\
&=&  \frac{1}{8\pi}\left[ \log \frac{\Lambda^2+m^2}{m^2} +e^{2m^2t}\left\{{\rm Ei}(-2t m^2) - {\rm Ei}(-2t (\Lambda^2+m^2))\right\}\right] 
\eeqa
and ${\rm E_i(x)}$ is the exponential integral function defined by
\beqa
{\rm Ei}(-x) =\int dx\, \frac{e^{-x}}{x} .
\eeqa


\section{Finiteness of the two point function}
\label{sec:finiteness}
In this section, we show the finiteness of the two point function in terms of the gradient field $\phi^\alpha$ non-perturbatively at the leading order of large $N$ expansion, without the field renormalization.

Since the leading behavior  of $\langle \varphi^a\varphi^b (\varphi\varphi)^n\rangle_c$ is $N^{-(2n+1)+n} = N^{-(n+1)}$,  the leading contribution to the two point function is simply given by
\beqa
\langle \phi^a(t_1, p_1) \phi^b(t_2, p_2) \rangle &=& f(t_1) f(t_2)e^{-p_1^2 t_1}e^{-p_2^2 t_2}\langle \varphi^a(p_1) \varphi^b(p_2) \rangle
\nn \\
&=& \frac{f(t_1)f(t_2) \lambda}{N}\delta^{ab}\hat{\delta}(p_1+p_2)\frac{e^{-p_1^2(t_1+t_2)}}{p_1^2+m^2} .
\eeqa 
Using
\beqa
f(t) &=&  \sqrt{\frac{\log(1+\Lambda^2/m^2)}{{\rm Ei}(-2t(\Lambda^2+m^2))-{\rm Ei(-2tm^2)}}} e^{-m^2 t},
\eeqa
which implies
\beqa
\lim_{\Lambda\rightarrow\infty} \lambda f(t_1) f(t_2) &=& 4\pi \frac{e^{-m^2(t_1+t_2)}}{ \sqrt{- {\rm Ei} (-2t_1 m^2)}\sqrt{- {\rm Ei} (-2t_2 m^2)}},
\eeqa
two point function is shown to be finite as
\beqa
\langle \phi^a(t_1, p_1) \phi^b(t_2, p_2) \rangle&=& 
\frac{4\pi e^{-(p_1^2+m^2)(t_1+t_2)}\delta^{ab}\hat{\delta}(p_1+p_2)}{ N\sqrt{- {\rm Ei} (-2t_1 m^2)}\sqrt{- {\rm Ei} (-2t_2 m^2)}} \frac{1}{p_1^2+m^2}
\label{eq:grad-2pt}
\eeqa
as long as $t_1 t_2 \not= 0 $, without renormalization factor for the field $\phi$. This is the main result of this paper.

At small $t_1, t_2$, we have
\beqa
\langle \phi^a(t_1, p_1) \phi^b(t_2, p_2) \rangle &=& 
\frac{4\pi \delta^{ab}\hat{\delta}(p_1+p_2)}{ N\sqrt{- \log t_1}\sqrt{- \log t_2}} \frac{1}{p_1^2+m^2},
\eeqa
which diverges as $ 1/\sqrt{\log t_1\log t_2}$ in the $t_1, t_2 \rightarrow 0$ limit.

\section{Applications}
\label{sec:application}
One of the applications of the gradient flow is the new definition of the running coupling constant.
Let us see what is the case for the two dimensional O(N) nonlinear sigma model at large $N$.
A scheme for  running coupling constant can be defined by the energy density.
Consider the vacuum expectation value of the energy density 
\begin{eqnarray}
E(\phi(t,x))  \equiv \langle \frac{1}{2}\sum_{\alpha=1}^N(\partial_\mu\phi^\alpha)^2 (t,x) \rangle .
\end{eqnarray}
At leading order in perturbation theory, it can be evaluated as 
\begin{eqnarray}
E_{\rm leading P.T. } = \frac{\lambda}{2} \int \frac{d^2 p}{(2\pi)^2} \frac {p^2}{p^2}e^{-2p^2t}
= \frac{\lambda}{16\pi t}.
\end{eqnarray}
Then, one can define the non-perturbative running coupling constant $\lambda_R(\mu)$ with the renormalization scale 
$\mu=8 \pi / t$ as 
\begin{eqnarray}
E_{\rm N.P.} = \frac{\lambda_R(\mu)}{16\pi t}.
\label{eq:E_NP}
\end{eqnarray}
From our non-perturbative result, the left hand side of eq. (\ref{eq:E_NP}) is evaluated as
\begin{eqnarray}
E_{\rm N.P. } = \frac{\lambda f(t)^2}{2} \int \frac{d^2p}{(2\pi)^2} \frac{p^2}{p^2+m^2} e^{-2p^2t}.
\label{eq:E_NP_result}
\end{eqnarray}
Combing eqs. (\ref{eq:E_NP}) and (\ref{eq:E_NP_result}), one finds 
\begin{eqnarray}
\lambda_R(\mu) &=&\frac{1}{\int \frac{d^2p}{(2\pi)^2} \frac{e^{-2p^2t}}{p^2+m^2} }\int_0^\infty dx \frac{xe^{-x}}{x+2m^2t} 
\nonumber\\
&\approx& \frac{1}{\int \frac{d^2p}{(2\pi)^2} \frac{e^{-2p^2t}}{p^2+m^2} } \left[ 1 + O(m^2 t) \right]
\end{eqnarray}
Recalling the gap equation (\ref{eq:Gap}), the renormalized coupling reduces to the bare coupling at $t=0$, 
and for finite $t$ it is  a UV finite coupling at the fully nonpertubative level defined by a momentum integral in which 
the cutoff $\Lambda$ is replaced with an effective cutoff of order $\sqrt{1/t}$. 

\section{Connected four point function}
\label{sec:discussions}
So far we have calculated the two point function of the gradient flow field non-perturbatively at the leading order of the large $N$ expansion, by solving the gradient flow equation necessary for the calculation. We then have shown the finiteness  of the two point function without renormalization, as is given in \eref{eq:grad-2pt}.
In this section, we extend our analysis to the connected four point function, which requires the next to leading order terms in the $1/N$ expansion.

To calculate the connected four point function, we have to determine  $X_3^a(\varphi,p,t)$ in \eref{eq:sol} by the gradient flow equation, which leads to
\beqa
X_3^i (\varphi,p,t) &=&   \lambda \int_p^3  (p_2\cdot p_3) \varphi^i(p_1)\  \varphi(p_2)\cdot  \varphi(p_3) X(p_1,p_2,p_3, t)
\eeqa
where $X(p_1,p_2,p_3,t)$ is given in \eref{eq:defX} and $\varphi\cdot\varphi =\sum_{a=1}^{N-1}(\varphi^a)^2$.
See Appendix \ref{app:sol-4pt} for the detail of the derivation.

We then consider  the power counting in the large $N$ expansion for the four point function of the gradient flow fields.
There are three types of contributions for $\varphi$ correlation functions.
\begin{itemize}
\item $\langle \varphi^a\varphi^b \varphi^c\varphi^d (\varphi\varphi)^n\rangle_c$:
the leading behavior is $N^{-(2n+3)+n} = N^{-(n+3)}$.
\item $\langle \varphi^a\varphi^b (\varphi\varphi)^n\rangle_c$:
the leading behavior is $N^{-(2n+1)+n} = N^{-(n+1)}$.
\item $\langle (\varphi\varphi)^n\rangle_c$:
the leading behavior is $N^{-(2n-1)+n} = N^{-(n-1)}$.
\end{itemize}
Therefore, contributions which have the leading behavior of the connected four point function, $N^{-3}$, are the following three types.
\begin{enumerate}
\item $\langle \varphi^a\varphi^b \varphi^c\varphi^d \rangle_c$
\item $\langle \varphi^a\varphi^b \rangle_c \langle \varphi^c\varphi^d (\varphi\varphi)\rangle_c$ where
$(\varphi\varphi)$ comes from $\varphi^a$ or $\varphi^b$.
\item $\langle \varphi^a\varphi^b\rangle_c  \langle \varphi^c\varphi^d\rangle_c\langle (\varphi\varphi)^2\rangle_c$ where one $(\varphi\varphi)$ comes from $\varphi^a$ or $\varphi^b$ and the other from $\varphi^c$ or $\varphi^d$.
\end{enumerate}

Now we calculate the connected part of four point function for the gradient flow fields as
\beqa
&&\langle \phi^{a_1}(p_1,t_1) \phi^{a_2}(p_2,t_2) \phi^{a_3}(p_3,t_3) \phi^{a_4}(p_4,t_4) \rangle_c \nn \\
&=&\prod_{i=1}^4 f(t_i) e^{-p_i^2 t_i} \Bigl[
\langle \varphi^{a_1}(p_1)  \varphi^{a_2}(p_2)  \varphi^{a_3}(p_3)  \varphi^{a_4}(p_4) \rangle_c \nn \\
&+& \left\{ \langle :X_3^{a_1}(\varphi,p_1,t_1): \varphi^{a_2}(p_2)  \varphi^{a_3}(p_3)  \varphi^{a_4}(p_4) \rangle_c 
+  \mbox{3 permutations} \right\} \nn \\
&+& \left\{ \langle :X_3^{a_1}(\varphi,p_1,t_1):  :X_3^{a_2}(\varphi,p_2,t_2):  \varphi^{a_3}(p_3)  \varphi^{a_4}(p_4) \rangle_c 
+  \mbox{5 permutations} \right\}
 \Bigr] \nn \\
\eeqa
where the first term
\beqa
\langle \varphi^{a_1}(p_1)  \varphi^{a_2}(p_2)  \varphi^{a_3}(p_3)  \varphi^{a_4}(p_4) \rangle_c &=& 
G_{a_1a_2a_3a_4}^{(4)}(p_1,p_2,p_3,p_4) 
\eeqa
is given in eq.~(\ref{eq:4-pt}) of Appendix \ref{app:4-pt}.
The second term can be evaluated as
\beqa
&&\langle :X_3^{a_1}(\varphi,p_1,t_1): \varphi^{a_2}(p_2)  \varphi^{a_3}(p_3)  \varphi^{a_4}(p_4) \rangle_c 
\equiv GX^{a_1}_{a_2a_3a_4}(p_1,p_2,p_3,p_4,t_1)
\nn \\
&=&
\hat\delta\left(p_{1234}\right) \frac{\lambda^4}{N^3} \prod_{i\not= 1} \frac{1}{p_i^2+m^2}  
 \Bigl[ \delta_{a_1a_2}\delta_{a_3a_4} \Bigl\{2 X(p_2,p_3,p_4,t_1) (p_3\cdot p_4) \nn \\
 &-& G(p_{34}) T(p_2,p_{34}, t_1)\Bigr\}
 + (2\leftrightarrow 3) +  (2\leftrightarrow 4)\Bigr]
\eeqa
where
\beqa
T(p_2,p_{34},t_1) &=& \int \prod_{i=1}^2 \frac{d^2 q_i }{(2\pi)^2} \hat\delta(q_{12}+p_{34})
\frac{(q_1\cdot q_2) X_3(p_2,q_1,q_2,t_1)}{(q_1^2+m^2)(q_2^2+m^2)}
\eeqa
with $p_{ij}=p_i+p_j$, $p_{ijkl}=p_{ij}+p_{kl}$, 
while the third term is given by
\beqa
&&\langle :X_3^{a_1}(\varphi,p_1,t_1):  :X_3^{a_2}(\varphi,p_2,t_2):  \varphi^{a_3}(p_3)  \varphi^{a_4}(p_4) \rangle_c \equiv GY^{a_1a_2}_{a_3a_4}(p_1,p_2,p_3,p_4,t_1,t_2)  \nn \\
&=&-\delta_{a_1a_3}\delta_{a_2a_4}\frac{\lambda^6}{N^3}  \hat\delta(p_{1234})
\prod_{i=3}^4\frac{1}{p_i^2+m^2}
T(p_3,p_{13},t_1)T(p_4,p_{24},t_2)G(p_{13})\nn \\
&+& ( 3\leftrightarrow 4) .
\eeqa

We finally obtain
\beqa
&&\langle \varphi^{a_1}(t_1,p_1) \varphi^{a_2}(t_2,p_1) \varphi^{a_3}(t_3,p_3) \varphi^{a_4}(t_4,p_4) \rangle_c \nn \\
&=&\prod_{i=1}^4 f(t_i) e^{-p_i^2 t_i} \Bigl[ G_{a_1a_2a_3a_4}^{(4)}(p_1,p_2,p_3,p_4) +GA^{(4)}_{a_1a_2a_3a_4}(p_1,t_1,p_2,t_2,p_3,t_3,p_4,t_4) \nn \\
&&+ GB^{(4)}_{a_1a_2a_3a_4}(p_1,t_1,p_2,t_2,p_3,t_3,p_4,t_4)
\Bigr]
\label{eq:grad-4pt_full}
\eeqa
where
\beqa
GA^{(4)}_{a_1a_2a_3a_4}(p_1,t_1,p_2,t_2,p_3,t_3,p_4,t_4) &\equiv& GX^{a_1}_{a_2a_3a_4}(p_1,p_2,p_3,p_4,t_1) \nn \\
&+& \mbox{3 permutations} , \\
GB^{(4)}_{a_1a_2a_3a_4}(p_1,t_1,p_2,t_2,p_3,t_3,p_4,t_4) &\equiv& GY^{a_1a_2}_{a_3a_4}(p_1,p_2,p_3,p_4,t_1,t_2) \nn \\
&+& \mbox{5 permutations}  .
\eeqa

It is important  to investigate whether \eref{eq:grad-4pt_full} is finite or not.
We however leave this investigation to future studies since analysis is so involved due to the complicated structure of  \eref{eq:grad-4pt_full}.

\section{Summary}\label{sec:summary}
In this paper, we studied the gradient flow in two dimensional $O(N)$ nonlinear sigma model.
Introducing the normal ordering technique and expanding in powers of the original bare field $\varphi^a(x)$, 
we have shown that the solution of the gradient flow equation $\phi^a(t,x)$ at flow time $t$ can be parameterized 
by the functions $X_{2n+1}$ $(n=0,1, \cdots)$, which correspond to the contributions from connected diagrams 
which scales  $O(1/N^{2n+1})$  at leading order in $1/N$ expansion. Since the differential equation 
for  $X_{2n+1}$ involves only lower order coefficients $X_{2m+1} (m=0,1,\cdots, n)$,  the solutions for the coefficient 
functions can be obtained recursively. 

In this work, we have found analytic solutions for $X_1$ and $X_3$.  
Using the explicit form of $X_1$, we have shown that the two point function 
$\langle \phi^a(t,\mathbf{x}) \phi^b(s,\mathbf{y})\rangle $ is finite.
Here two remarks are in order. At leading order in $1/N$,  any $2n$ point  functions
$\langle \prod_{i=1}^{2n}  \phi^a(t_i,\mathbf{x_i})\rangle $ can be factorized into the 
product of two point functions. 
Therefore any correlation functions including composite operators  
constructed from the fields at finite flow time are shown  to be finite non-perturbatively in the large $N$ limit.
In this sense, the field $\phi$ and its composite operators are automatically renormalized at finite flow time
in the large $N$ limit.

Further  question would be 
whether the connected contribution to the correlation functions is finite or not.
The simplest nontrivial example to this question is the finiteness of the 
connected contribution to the four point function.
In principle using our solution $X_3$ we can examine whether the connected four point function 
is finite or not. However, our solution is only a formal one which involves a function in terms of the integral operator $F$. Due to 
its complicated structure we could not yet succeeded in extracting out the ultraviolet divergences 
so that whether the connected four point function is finite or not is yet to be seen.

\appendix


\section{Connected four point function in the two dimensional O(N) nonlinear sigma model}
\label{app:4-pt}
The connected four point function at the leading order of the large $N$ expansion can be calculated as
\beqa
\langle \varphi^{a_1}(x_1) \varphi^{a_2}(x_2) \varphi^{a_3}(x_3) \varphi^{a_4}(x_4)\rangle_c &=&
\frac{\delta}{\delta J^{a_1}(x_1)} \frac{\delta}{\delta J^{a_2}(x_2)}\frac{\delta}{\delta J^{a_3}(x_3)}\frac{\delta}{\delta J^{a_4}(x_4)}
F_4 J^4
\eeqa
with $F_4$ in \eref{eq:F4}, which  leads to
\beqa
&=& - 4\frac{\lambda^4}{N^3} \left[ \delta_{a_1a_2}\delta_{a_3a_4} K_0^{-1}(x_1,x) K_0^{-1}(x_2,x) D_0^{-1}(x,y) K_0^{-1}(x_3-y)  K_0^{-1}(x_4-y)
\right. \nn \\
&&\left. + (2\leftrightarrow 3) + (2\leftrightarrow 4) \right] .
\eeqa
In the momentum space, we obtain
\beqa
&&G_{a_1a_2a_3a_4}^{(4)}(p_1,p_2,p_3,p_4) \equiv \langle \varphi^{a_1}(p_1) \varphi^{a_2}(p_2) \varphi^{a_3}(p_3) \varphi^{a_4}(p_4)\rangle_c \nn \\
&=&- \hat\delta \left(p_{1234}\right) 
4\frac{\lambda^4}{N^3} \prod_{j=1}^4 K_0^{-1}(p_j) \left[ \delta_{a_1a_2}\delta_{a_2a_3} D_0^{-1}(p_{12}) +  (2\leftrightarrow 3) + (2\leftrightarrow 4) \right]
\eeqa
where 
\beqa
D_0(p)&=& 2\lambda^2 \int\frac{d^2 q}{(2\pi)^2} K_0^{-1}(q) K_0^{-1}(p-q) \nn \\
&=& \frac{\lambda^2}{\pi \vert p\vert\sqrt{p^2+4m^2}} \log \frac{\sqrt{p^2+4m^2}+\vert p\vert}{\sqrt{p^2+4m^2}-\vert p\vert}.
\eeqa
We then  finally obtain
\beqa
G_{a_1a_2a_3a_4}^{(4)}(p_1,p_2,p_3,p_4) &=& -\hat\delta (p_{1234})\frac{ \lambda^2}{N^3}
\prod_{j=1}^4 \frac{1}{p_j^2 + m^2}\nn \\
&\times& \Bigl\{\delta_{a_1a_2}\delta_{a_3a_4}G (p_{12}) +  (2\leftrightarrow 3) + (2\leftrightarrow 4) \Bigr\}
\label{eq:4-pt}
\eeqa
where
\beqa
G(p) &=& 2\pi \frac{\vert p\vert \sqrt{p^2+4m^2}}{ \log \dfrac{\sqrt{p^2+4m^2}+\vert p\vert}{2m}} .
\eeqa
The connected four point function is $O(1/N^3)$, which is $O(1/N)$ smaller than the $O(1/N^2)$ disconnected contribution.
 
 \section{Solution to the gradient flow equation for $X_3$}
\label{app:sol-4pt}
Taking ${\cal O}_3$ as the order $\varphi^3$ operator, we evaluate $L^a$ and $R^a$ at the leading order of the large $N$ expansion as
\beqa
\langle L^a(t,p) {\cal O}_3 \rangle &=& e^{-p^2 t}\left[ \dot f(t) \langle : X_3^a (\varphi,p,t): {\cal O}_3 \rangle
+ f(t)  \langle : \dot X_3^a (\varphi,p,t): {\cal O}_3 \rangle
\right] \\
\langle R^a(t,p) {\cal O}_3 \rangle &=& e^{-p^2 t} f^3(t) I(t) \langle  : X_3^a (\varphi,p,t): {\cal O}_3 \rangle
-f^3(t) \int_p^3 e^{-\sum_{j=1}^3   p_j^2 t}\nn \\
&\times&( p_2\cdot p_3) \langle :\varphi^a(p_1)\  \varphi(p_2)\cdot  \varphi(p_3):
{\cal O}_3\rangle -2f^3(t)  \int_p^3 e^{-\sum_{j=1}^3   p_j^2 t} \nn \\
&\times& ( p_2\cdot p_3) \langle \varphi^a(p_1) : X_3^b(\varphi,p_2,t):  \varphi^b(p_3)
{\cal O}_3\rangle
+O(1/N) ,
\eeqa
where in the last term,  $\varphi^b$ in $X_3^b(\varphi,p_2,t)$ is contracted with $\varphi^b(p_3)$.
Since $\dot f(t) = f^3(t) I(t)$, the first terms in both sides agree. 
Therefore, the equation we have to solve at the leading order becomes
\beqa
  \langle : \dot X_3^a (\varphi,p,t): {\cal O}_3 \rangle
&=& 
- f^2(t) e^{p^2 t}  \int_p^3 e^{-\sum_{j=1}^3   p_j^2 t}(p_2\cdot p_3)
\left[ \langle :\varphi^a(p_1)\  \varphi(p_2)\cdot  \varphi(p_3): {\cal O}_3\rangle \right. \nn \\
&+& \left. 2 \langle \varphi^a(p_1) : X_3^b(\varphi,p_2,t):  \varphi^b(p_3)
{\cal O}_3\rangle \right].
\label{eq:order3}
\eeqa

Since the above equation is difficult to solve directly, we introduce the expansion in $\lambda$ as
\beqa
X_3^a (\varphi,p,t) &=& \sum_{n=0}^\infty \lambda^n X_{3,n}^a (\varphi,p,t) ,
\label{eq:expand}
\eeqa
where $X_{3,n}$ is independent on $\lambda$. From \eref{eq:order3}, we can easily obtain
\beqa
X_{3,0}^a (\varphi,p,t) &=& - \int_p^3 
  (p_2\cdot p_3) \varphi^a(p_1)\  \varphi(p_2)\cdot  \varphi(p_3)  \int_0^t d\, s f^2(s) e^{p^2 s}e^{-(p_1^2+p_2^2+p_3^2)s}, 
\eeqa
which, after a little algebra, leads to
\beqa
X_{3,1}^a (\varphi,p,t) &=& - 2\int_p^3 (p_2\cdot p_3) \varphi^a(p_1)\  \varphi(p_2)\cdot  \varphi(p_3)  \int_0^{t} d\,s_1  f^2(s_1) e^{(p^2-p_1^2)s_1}\nn \\
&\times& \int_0^{s_1} d\, s_0\, 
J_0(p_{23}, s_1,s_0)\, e^{-(p_2^2+p_3^2)s_0}
\label{eq:sol1}
\eeqa
where
\beqa
J_0(p_{23},t,s) &=& - f^2(s) 
 \left(\prod_{j=2}^3 \int \frac{d^2 q_j}{(2\pi)^2}e^{-q_j^2 t}\right) e^{(q_2^2-q_3^2)s} \frac{ q_2\cdot q_3 } {q_3^2+m^2}\hat\delta(q_{23}-p_{23})
\eeqa
It is then not so difficult to guess the solution for a general $n$ as
\beqa
X_{3,n}^a  (\varphi,p,t) &=&-\int_p^3
(p_2\cdot p_3) \varphi^a(p_1)\  \varphi(p_2)\cdot  \varphi(p_3) 
 \int_0^{t} d\,s_{n}  f^2(s_{n}) e^{(p^2-p_1^2)s_{n}}\nn \\
&\times& \left(   \prod_{i=n-1}^0 \int_0^{s_{i+1}} d\,s_i\, 2 J_0(p_{23},s_{i+1}, s_i)  \right)  e^{-(p_2^2+p_3^2)s_0}, 
\label{eq:soln}    
\eeqa
which can be proven by the mathematical induction as follows.
The solution for $n=1$ is correct by \eref{eq:sol1}.
If \eref{eq:soln} is correct for $n= k$, \eref{eq:order3} gives
\beqa
\dot X_{3,k+1}^a  (\varphi,p,t) &=& - f^2(t)e^{p^2 t}
\int_p^3 e^{-p_1^2 t}  (p_2\cdot p_3) \varphi^a(p_1)\  \varphi(p_2)\cdot  \varphi(p_3) \nn \\
&\times& \left( \prod_{i=k}^0 \int_0^{s_{i+1}} d\,s_i\, 2 J_0(p_{23},s_{i+1}, s_i)  \right)
 e^{-(p_2^2+p_3^2)s_0}
 \eeqa 
with $s_{k+1}=t$. By integrating the above equation in $t$, we show that \eref{eq:soln} is correct for $n=k+1$.
This completes the proof.

We now introduce the integral operator $F(p_{23})$ and a function $H(p_2^2+p_3^2)$ as
\beqa
H(p_2^2+p_3^2)[t] &\equiv& e^{-(p_2^2+ p_3^2)t}, \\
\left[ F(p_{23}) H(p_2^2+p_3^2)\right] [t] &\equiv& 
\int_0^\infty d\,s\,   \Theta(t-s) J_0(p_{23},t,s) H(p_2^2+p_3^2)[s].
\eeqa 
Using these notations, $X_{3,n}$ for all $n$ can be expressed as
\beqa
X_{3,n}^a  (\varphi,p,t) &=& -\int_p^3 (p_2\cdot p_3) \varphi^a(p_1)\  \varphi(p_2)\cdot  \varphi(p_3)
\nn \\ &\times&  
  \int_0^{t} d\,s\,  f^2(s) e^{(p^2-p_1^2)s} 2^n \left[  F^n(p_{23}) H(p_2^2+p_3^2)\right][s] .
\eeqa
Combining this with \eref{eq:expand}, we finally obtain
\beqa
X_3^a (\varphi,p,t) &=&  -\int_p^3  (p_2\cdot p_3) \varphi^a(p_1)\  \varphi(p_2)\cdot  \varphi(p_3)\nn \\
 &\times&    
 \int_0^{t} d\,s\,  f^2(s) e^{(p^2-p_1^2)s} \left[ \frac{1}{1-2 \lambda F(p_{23})} H(p_2^2+p_3^2)\right][s] 
 \nn \\
 &\equiv&  \int_p^3  (p_2\cdot p_3) \varphi^a(p_1)\  \varphi(p_2)\cdot  \varphi(p_3) X(p_1,p_2,p_3, t)
\eeqa
where
\beqa
X(p_1,p_2,p_3,t) &=& - \int_0^{t} d\,s\,  f^2(s) e^{(p^2-p_1^2)s} \left[ \frac{1}{1-2 \lambda F(p_{23})} H(p_2^2+p_3^2)\right][s] .
\label{eq:defX}
\eeqa

\section{An alternative way to solve the flow equation}
\label{app:alternative}
In this appendix, we present an alternatively way to solve the flow equation. 
Instead of solving the field at flow time $t$ in terms of bare fields, we derive 
the differential equation on the correlation function for the fields at finite time $t$.
\subsection{Schwinger-Dyson equation}
General correlation function 
$\langle \phi^a(t,x) \mathcal{O} \rangle$, where $\mathcal{O}$ is 
an arbitrary operator constructed from $\phi$,  satisfies the following differential 
equation.
\begin{eqnarray}
\frac{d}{dt}\langle \phi^a(t,x) \mathcal{O} \rangle 
&=& \langle \square\phi^a(t,x) \mathcal{O}\rangle 
+\langle \phi^a(t,x)(\partial_\mu\vec{\phi}(t,x))^2\mathcal{O}\rangle
\nonumber\\
&&+\frac{1}{4}\sum_{n=0}^\infty
\langle \phi^a(t,x)(\partial_\mu \vec{\phi}^2(t,x))^2 (\vec{\phi}(t,x))^{2n} \mathcal{O}\rangle.
\end{eqnarray}
In momentum representation the differential equation reads
\begin{eqnarray}
\frac{d}{dt}\langle \phi^a(t, p) \mathcal{O}\rangle
&=&-p^2 \langle \phi^a(t, p) \mathcal{O}\rangle
 -\int_p^3 (p_2\cdot p_3) \langle \phi^a(t, p_1)\vec{\phi}(t,p_2) \cdot\vec{\phi}(t,p_3) 
 \mathcal{O}\rangle
\nonumber\\
 &-&\sum_{n=0}^\infty
\int_p^{2n+5}\frac{(p_2+p_3)\cdot (p_4+p_5) }{4}
\nonumber\\
&& \times \langle \phi^a(t,p_1)
\prod_{j=1}^2(\vec{\phi}(t,p_{2j})\cdot\vec{\phi}(t,p_{2j+1}) )
\prod_{l=0}^n(\vec{\phi}(t,p_{2l+6})\cdot \vec{\phi}(t,p_{2l+7}) )
\mathcal{O}\rangle.
 \nonumber\\
\end{eqnarray}


\subsection{Leading Contribution for two point function}
Let us consider two point function.
Setting $t=t_a$, $p=p_a$ and choosing $\mathcal{O}=\phi(t_b,p_b)$, 
we obtain the differential equation for the two point function as
\begin{eqnarray}
&&\frac{d}{dt_a}\langle \phi^a(t_a, p_a) \phi(t_b,p_b)\rangle
=-p_a^2 \langle \phi^a(t_a, p) \phi(t_b,p_b)\rangle
\nn \\
&-&\int _{p_a}^3(p_2\cdot p_3) \langle \phi^a(t_a, p_1)
\vec{\phi}(t_a,p_2) \cdot\vec{\phi}(t_a,p_3) 
 \phi^b(t_b,p_b)\rangle
 -\sum_{n=0}^\infty
\int _{p_a}^{2n+5}\frac{(p_2+p_3)\cdot (p_4+ p_5)}{4} 
\nonumber\\
&& \times \langle \phi^a(t_a,p_1)
\prod_{j=1}^2(\vec{\phi}(t_a,p_{2j})\cdot\vec{\phi}(t_a,p_{2j+1}) )
\prod_{l=0}^n(\vec{\phi}(t_a,p_{2l+6})\cdot \vec{\phi}(t_a,p_{2l+7}) )
\phi(t_b,p_b)\rangle.
 \nonumber\\
 \label{eq:Deq-2pt}
\end{eqnarray}

Let us now consider the leading order contribution at large $N$.
In Section~\ref{sec:model}, we have shown that the leading order contribution to the two point function is $O(1/N)$.
Therefore, we should only consider $O(1/N)$ contribution on the right hand side of eq. (\ref{eq:Deq-2pt}).
The four point function in the second term on the right hand side can be decomposed as 
\begin{eqnarray}
\langle \phi^a(t_a, p_1)\vec{\phi}(t_a,p_2) \cdot\vec{\phi}(t_a,p_3) &\phi^b&(t_b,p_b)\rangle
= \langle \phi^a(t_a, p_1)\vec{\phi}(t_a,p_2) \cdot\vec{\phi}(t_a,p_3) \phi^b(t_b,p_b)\rangle_c
\nonumber\\
&+&  \langle \phi^a(t_a, p_1)\vec{\phi}(t_a,p_2) \rangle \cdot \langle\vec{\phi}(t_a,p_3) \phi^b(t_b,p_b)\rangle
\nonumber\\
&+&  \langle \phi^a(t_a, p_1)\vec{\phi}(t_a,p_3) \rangle \cdot \langle\vec{\phi}(t_a,p_2) \phi^b(t_b,p_b)\rangle
\nonumber\\
&+&  \langle \phi^a(t_a, p_1) \phi^b(t_b,p_b)\rangle \langle\vec{\phi}(t_a,p_2)\cdot \vec{\phi}(t_a,p_3)\rangle, 
\label{eq:4pt}
\end{eqnarray}
where $\langle \dots \rangle_c$ denotes the connected parts.

In Section~\ref{sec:model}, we have also shown that the leading order connected parts in the 2n-point function is of $O(1/N^{2n-1})$. 
Dut to the O(N) symmetry, the two point function $\langle \phi^a \phi^b\rangle$ is proportional to $\delta^{ab}$ 
the four point function $\langle \phi^a \phi^b \phi^c \phi^d\rangle$ 
can be decomposed into the sum of three functions which are proportional to 
$\delta^{ab}\delta^{cd}$, $\delta^{ac}\delta^{bd}$, $\delta^{ad}\delta^{bc}$, respectively.
From this fact, one can see that the first, the second and the third terms on the right hand side of eq. (\ref{eq:4pt}) are $O(1/N^2)$, 
whereas the fourth term is $O(1/N)$.

Similar argument can be applied to the third term of eq. (\ref{eq:Deq-2pt}).
For example, six point function in the term with $n=0$ can be decomposed as follows
\begin{eqnarray}
&&\langle \phi^a(t_a,p_1)\vec{\phi}(t_a,p_2)\cdot\vec{\phi}(t,p_3) \vec{\phi}(t_a,p_4)\cdot\vec{\phi}(t,p_5)  \phi^b(t_b,p_b)\rangle
\nonumber\\
&=&\langle \phi^a(t_a,p_1)\vec{\phi}(t_a,p_2)\cdot\vec{\phi}(t,p_3) \vec{\phi}(t_a,p_4)\cdot\vec{\phi}(t,p_5)  \phi^b(t_b,p_b)\rangle_c
\nonumber\\
&+& \langle \phi^a(t_a,p_1)\phi^b(t_b,p_b)\rangle
\langle \vec{\phi}(t_a,p_2)\cdot\vec{\phi}(t,p_3) \vec{\phi}(t_a,p_4)\cdot\vec{\phi}(t,p_5) \rangle_c
+ \mbox{ other  2pt $\times$ 4pt}
\nonumber\\
&  + &\langle \phi^a(t_a,p_1)\phi^b(t_b,p_b)\rangle
\langle \vec{\phi}(t_a,p_2)\cdot\vec{\phi}(t,p_3) \rangle \langle \vec{\phi}(t_a,p_4)\cdot\vec{\phi}(t,p_5) \rangle
+ \mbox{ other products of 2pt}.
\nonumber\\
\label{eq:6pt}
\end{eqnarray}
The first and second terms on the right hand side of eq. (\ref{eq:6pt}) are $O(1/N^3)$ , $O(1/N^2)$ or higher.
In the third term only the first contribution gives $O(1/N)$ and "other products of 2pt" give only higher order contributions.
It is found that  the $O(1/N)$ in the third term contains a factor $(p_2+p_3)\cdot(p_4+p_5)$. Due to the momentum conservation
for the two point functions, they only give vanishing contribution. From similar observation, one finds that there is no contribution from the third term of 
\eref{eq:Deq-2pt}. 

From this consideration, one finally finds that the gradient flow equation at the leading order reduces to 
\begin{eqnarray}
&&\frac{d}{dt_a}\langle \phi^a(t_a, p_a) \phi^b(t_b,p_b)\rangle
\nonumber\\
&=&-p_a^2 \langle \phi^a(t_a, p_a) \phi^b(t_b,p_b)\rangle
-\int_{p_a}^{3}(p_2\cdot p_3) 
  \langle \phi^a(t_a, p_1)\phi^b(t_b,p_b)\rangle
\langle \vec{\phi}(t_a,p_2) \cdot\vec{\phi}(t_a,p_3) \rangle .
 \nonumber\\
  \label{eq:Deq-2pt-final}
\end{eqnarray}
This means that at leading order, one obtains a closed equation for the two point function. 

From momentum conservation and O(N) symmetry, the two point function takes the form
\begin{eqnarray}
\langle \phi^a(t_a,p_a)  \phi^b(t_b,p_b) \rangle = \hat{\delta}(p_a+p_b)\delta^{ab} G(t_a,t_b,p_a^2).
\end{eqnarray}
Substituting this into eq. (\ref{eq:Deq-2pt-final}), we obtain
\begin{eqnarray}
\frac{d}{dt_a}G(t_a,t_b,p_a^2)
= \left[-p_a^2+N \int \frac{d^2 q}{(2\pi)^2} q^2 G(t_a,t_a,q^2) \right]G(t_a,t_b,p_a^2)
\label{eq:flow-for-Gp}
\end{eqnarray}


\subsection{Exact solution of two point function at large N}
We employ the following ansatz for the two point function
\begin{eqnarray}
G(t_a,t_b,p_a^2)= f(t_a) f(t_b)    e^{-p_a^2 (t_a+t_b)} \frac{\lambda}{N(p_a^2+m^2)}
\label{eq:ansatz}
\end{eqnarray}
where $f(t)$ is some function of $t$.  In order to reproduce the propagator at $t=0$, $f(t)$ must 
satifsy the initial condition $f(0)=1$.

Substituing eq. (\ref{eq:ansatz})  into eq. (\ref{eq:flow-for-Gp}), one finds that
\begin{eqnarray}
\frac{df(t)}{dt} =- \lambda \frac{1}{2}f^3(t)  \frac{dJ_0(t)}{dt}
\label{eq:f_eq}
\end{eqnarray}
where 
\begin{eqnarray}
J_0(t) = \int \frac{d^2 q}{(2\pi)^2} \frac{\exp(-2q^2 t) }{q^2+m^2}.
\end{eqnarray}
Note that $J_0(t)$ is finite at finite $t$ owing to the suppression factor $\exp(-2q^2 t)$ 
in the momentum integration, while at $t=0$ it is logarithmically divergent.

Solving eq. (\ref{eq:f_eq}), one obtains the following solution for $f(t)$
\begin{eqnarray}
f(t) = \left[ 1 +\lambda(J_0(t)-J_0(0)) \right]^{-1/2}.
\end{eqnarray}
Using the Gap equation $\lambda J_0(0)=1$, $f(t)$ is determined as
\begin{eqnarray}
f(t) = \left[ \lambda J_0(t) \right]^{-1/2}.
\end{eqnarray}
Therefore the two point function is given as
\begin{eqnarray}
G(t_a,t_b,p_a^2)= [J_0(t_a) J_0(t_b)]^{-1/2}   e^{-p_a^2 (t_a+t_b)} \frac{1}{N(p_a^2+m^2)}.
\label{eq:2pt-final}
\end{eqnarray}
One can easily see that the two point function at nonzero flow time $t$ is free from divergence.


\subsection{Leading contribution to the connected four point function}
We consider the connected four point function defined as
 \begin{eqnarray}
&&\langle \phi^a(t_a, p_a) \phi^b(t_b, p_b) \phi^c(t_c, p_c) \phi^d(t_d,p_d)\rangle_c
\equiv 
\langle \phi^a(t_a, p_a) \phi^b(t_b, p_b) \phi^c(t_c, p_c) \phi^d(t_d,p_d)\rangle
\nonumber\\
&- &\left[ \langle \phi^a(t_a, p_a) \phi^b(t_b, p_b)\rangle \langle \phi^c(t_c, p_c) \phi^d(t_d,p_d)\rangle
+ ( b\leftrightarrow c ) +  ( b\leftrightarrow d ) \right].
 \end{eqnarray}
Applying the gradient flow equation, one obtains the following differential equation for the four point function.
\begin{eqnarray}
&&(\frac{d}{d t_a}+p_a^2)<\phi^a(t_a, p_a)\prod_{e=b, c, d}\phi^e (t_e, p_e)>_c
\nonumber\\
 &=&-\int_{p_a}^{3}(p_2\cdot p_3)
 \times \left[ \langle \phi^a(t_a, p_1)
\vec{\phi}(t_a,p_2) \cdot\vec{\phi}(t_a,p_3) 
 \prod_{e=b,c,d}\phi^e(t_e,p_e)\rangle \right.
 \nonumber\\
&&\displaystyle{\left. 
- \left(\langle \phi^a(t_a, p_1)\vec{\phi}(t_a,p_2) \cdot\vec{\phi}(t_a,p_3) \phi^b(t_b,p_v)\rangle 
\langle \phi^c(t_c, p_c) \phi^d(t_d,p_d)\rangle
 + (b\leftrightarrow c) + (b\leftrightarrow d) \right)  \right]}
\nonumber\\
%
&-&\sum_{n=0}^\infty
\int_{p_a}^{2n+5}\frac{(p_2+p_3)\cdot (p_4+ p_5)}{4} 
\nonumber\\
&& \left[\langle \phi^a(t_a,p_1)
\prod_{j=1}^2(\vec{\phi}(t_a,p_{2j})\cdot\vec{\phi}(t,p_{2j+1}) )
\prod_{l=0}^n(\vec{\phi}(t_a,p_{2l+6})\cdot \vec{\phi}(t,p_{2l+7}) )
\prod_{e=b,c,d}\phi(t_e,p_e)\rangle \right.
 \nonumber\\
 && - \left(\langle \phi^a(t_a,p_1)
\prod_{j=1}^2(\vec{\phi}(t_a,p_{2j})\cdot\vec{\phi}(t,p_{2j+1}) )
\prod_{l=0}^n(\vec{\phi}(t_a,p_{2l+6})\cdot \vec{\phi}(t,p_{2l+7}) )
\phi(t_b,p_b)\rangle \langle \phi^c(t_c, p_c) \phi^d(t_d,p_d)\rangle
\right.
 \nonumber\\
 &&
\displaystyle{\left. \left. (b\leftrightarrow c) + (b\leftrightarrow d) \right) \right]}.
 \label{eq:Deq-4pt}
\end{eqnarray}
As shown in Section~\ref{sec:model}, the left hand side is $O(1/N^3)$.
What is the $O(1/N^3)$  contribution on the right hand side?
In the first term, there appear six point function, which can be decomposed into connected and 
disconnected contributions as
\begin{eqnarray}
&&  \langle \phi^a(t_a, p_1) \vec{\phi}(t_a,p_2) \cdot\vec{\phi}(t_a,p_3)  \prod_{e=b,c,d}\phi^e(t_e,p_e)\rangle
\nonumber\\
&=&  \langle \phi^a(t_a, p_1) \vec{\phi}(t_a,p_2) \cdot\vec{\phi}(t_a,p_3)  \prod_{e=b,c,d}\phi^e(t_e,p_e)\rangle_c
\nonumber\\
& &  +  \left( \langle \vec{\phi}(t_a,p_2) \cdot\vec{\phi}(t_a,p_3)  \phi^c(t_c,p_c) \phi^d(t_d,p_d) \rangle_c 
       \langle \phi^a(t_a,p_1)  \phi^b(t_b,p_b)\rangle  + ( b\leftrightarrow c  )  + ( b\leftrightarrow d  ) \right)
       \nonumber\\
& &  +  
  \langle \phi^a(t_a,p_1) \prod_{e=b,c,d} \phi^e(t_e,p_e) \rangle_c 
   \langle \vec{\phi}(t_a,p_2) \cdot\vec{\phi}(t_a,p_3)  \rangle 
       \nonumber\\
& &  +  \left( \langle \phi^a(t_a, p_1) \vec{\phi}(t_a,p_2) \cdot\vec{\phi}(t_a,p_3)  \phi^b(t_b,p_b)\rangle_c 
       \langle \phi^c(t_c,p_c)  \phi^d(t_d,p_d)\rangle  + ( b\leftrightarrow c  )  + ( b\leftrightarrow d  ) \right)
       \nonumber\\
& &   + (\mbox{ other connected 4pt } \times \mbox{ 2pt } )
\nonumber\\
& &  + 2  \left(  \langle \phi^a(t_a,p_1)\phi^b(t_b,p_b)\rangle 
\langle \vec{\phi}(t_a,p_2)  \phi^c(t_c,p_c)\rangle \cdot 
\langle \vec{\phi}(t_a,p_3)  \phi^d(t_d,p_d) \rangle
 + ( b\leftrightarrow c  )  + ( b\leftrightarrow d  ) \right)
 \nonumber\\
& &   + (\mbox{ other 2pt } \times \mbox{ 2pt }  \times \mbox{ 2pt } ).
\label{eq:6pt_vev}
\end{eqnarray}
In this decomposition, the first term on the right hand side of eq. (\ref{eq:6pt_vev}) 
is $O(1/N^4)$ and the second and third terms are $O(1/N^3)$.
The fourth term on the right hand side of eq. (\ref{eq:6pt_vev}) 
is also $O(1/N^3)$, but it is cancelled with the subtraction terms in eq. (\ref{eq:Deq-4pt}). 
The fifth term  on the right hand side of eq. (\ref{eq:6pt}) is 
$O(1/N^4)$ or higher since the O(N) invariant pair $\vec{\phi}(t_a,p_2)\cdot \vec{\phi}(t_a,p_3)$ 
is split into different correlation functions. Out of the product of three two point functions, the sixth term  of eq. (\ref{eq:6pt_vev}) 
is $O(1/N^3)$ whereas others (seventh term) are cancelled with the subtraction terms in eq. (\ref{eq:Deq-4pt}). 
One therefore finds
\begin{eqnarray}
&&  \langle \phi^a(t_a, p_1) \vec{\phi}(t_a,p_2) \cdot\vec{\phi}(t_a,p_3)  \prod_{e=b,c,d}\phi^e(t_e,p_e)\rangle
 - ( \mbox{ subtraction terms } )
\nonumber\\
&=& 
\left[\langle \vec{\phi}(t_a,p_2)\cdot \vec{\phi}(t_a,p_3) \phi^c(t_c,p_c) \phi^d(t_d,p_d) \rangle_c
f(t_a)f(t_b)\frac{\lambda \delta^{ab}e^{-p_b^2(t_a+t_b)}}{N(p_b^2+m^2)} \hat{\delta}(p_1+p_b)\right.
\nonumber\\
&&\left. + ( b\leftrightarrow c ) + ( b\leftrightarrow d ) \right]
\nonumber\\
&+ & 
\langle \vec{\phi}(t_a,p_1) \prod_{e=b,c,d} \phi^e(t_e,p_e) \rangle_c
f(t_a)^2\frac{\lambda e^{-2p_2^2 t_a}}{(p_2^2+m^2)} \hat{\delta}(p_2+p_3)
\nonumber\\
&+&
2\frac{\lambda^3}{N^3} f(t_a)^3 \prod_{e=b,c,d}\left[ f(t_e)\frac{e^{-p_e^3(t_a+t_e)}}{p_e^2+m^2}\right]
\nonumber\\
&& \times \left( \hat{\delta}(p_1+p_b)\hat{\delta}(p_2+p_c)\hat{\delta}(p_3+p_d)
 +(b\leftrightarrow c)  +(b\leftrightarrow d) \right).
\nonumber\\
\label{eq:6pt_leading}
\end{eqnarray}
What about the second term of eq. (\ref{eq:Deq-4pt})? After careful study of order counting and the use of momentum conservation 
similar to the case of two point function, one finds that there is no $O(1/N^3)$. 

Substituting eq. (\ref{eq:6pt_leading}) and using the fact that the the second term of eq. (\ref{eq:Deq-4pt}) does not give 
leading order contribution, one finds 
\begin{eqnarray}
&&\left(\frac{d}{dt_a}+p_a^2 -\frac{1}{f(t_a)}\frac{d f(t_a)}{dt_a} \right) \langle \phi^a(t_a,p_a) \prod_{e=b,c,d} \phi^e(t_e,p_e)\rangle_c
\nonumber\\
%
&=&- \left(\prod_{i=2,3}\int\frac{d^2 q_i}{(2\pi)^2}\right) \left[ \hat{\delta}(q_{23}-p_{ab})
(q_2\cdot q_3)\langle \vec{\phi}(t_a,q_2)\cdot \vec{\phi}(t_a,q_3) \phi^c(t_c,p_c) \phi^c(t_d,p_d) \rangle_c \right.
\nonumber\\
&&\hspace{3cm} \left. \times\frac{\lambda \delta^{ab}e^{-p_b^2(t_a+t_b)}}{N(p_b^2+m^2)} f(t_a)f(t_b) 
+ ( b\leftrightarrow c) + ( b\leftrightarrow d)\right]
\nonumber\\
&&-2\frac{\lambda^3}{N^3}f(t_a)^3 \left(\prod_{e=b,c,d}f(t_e) \frac{e^{-p_e^2(t_a+t_e)}}{p_e^2+m^2}\right)
 \hat{\delta}(p_{abcd}) \left( (p_c\cdot p_d) + ( p_b \cdot p_d ) + ( p_c \cdot p_b ) \right),
\nonumber\\
\end{eqnarray}
where $q_{23}=q_2+q_3$ and $p_{23}=p_2+p_3$ and $p_{abcd}=p_a+p_b+p_c+p_d$.
%

We can see that the gradient flow gives a closed equation also for the four point function.
Note that the coefficients of this differential equation are finite, since they are 
expressed by the combination of $\lambda$ times the product of two   $f(t)$'s  and $f(t) =\left[\lambda J(t)\right]^{-1/2}$ 
so that the bare coupling $\lambda$ dependence is explicitly cancelled. 
This means that the differential equation is consistent with the case that the connected four point function would be 
finite. 


\acknowledgments
The authors would like to thank Hiroshi Suzuki for discussions. 
This work was supported by Grant-in-Aid for JSPS Fellows Grant Number 25$\cdot$1336 and by the Grant-in-Aid of the Japanese Ministry of Education (Nos. 25287046,26400248), by MEXT SPIRE and JICFuS and by US DOE grant de-sc0011941.

\section*{Note added}
The recent paper\cite{Makino:2014cxa}, using the different method,  provides the two point function of the gradient flow fields in the same model, which  turns out to be  consistent with ours.









\providecommand{\href}[2]{#2}\begingroup\raggedright\endgroup

\end{document}